\documentclass[twocolumn]{aastex701}
\usepackage{amssymb}

\graphicspath{{./}{figures/}}

\usepackage{xcolor,color}
\usepackage[normalem]{ulem}
\usepackage{soul}
\usepackage{hyperref}
\usepackage{indentfirst}
\usepackage{bm}
\usepackage{gensymb}

\newcommand{\mgas}[0]{\ensuremath{M_{\rm gas}}}
\newcommand{\mdust}[0]{\ensuremath{M_{\rm dust}}}
\newcommand{\msun}[0]{\ensuremath{\mathrm{M}_{\odot}}}
\newcommand{\rc}[0]{\ensuremath{R_{\rm c}}}

\begin{document}

\title{Measuring the gas mass of the protoplanetary disk around HD~179218 with CO isotopologues}

\author[0009-0006-4575-633X]{Nina Weichmann}
\affiliation{Department of Astronomy, University of Wisconsin-Madison,
475 N Charter St, Madison, WI 53706}
\email{Nina.Weichmann@colorado.edu} 

\author[0000-0002-0661-7517]{Ke Zhang}
\affiliation{Department of Astronomy, University of Wisconsin-Madison,
475 N Charter St, Madison, WI 53706}
\email{ke.zhang@astro.wisc.edu} 

\author[0000-0002-8623-9703]{Leon Trapman}
\affiliation{Department of Astronomy, University of Wisconsin-Madison,
475 N Charter St, Madison, WI 53706}
\email{} 

\author[0000-0001-8500-2416]{Ao Zhang}
\affiliation{Department of Physics, Washington University in St. Louis,
ONE BROOKINGS DRIVE, St. Louis, MO 63130}
\email{}

\begin{abstract}
Carbon monoxide (CO) isotopologues are widely used to estimate protoplanetary disk gas masses, but their reliability is often compromised by CO freeze-out and subsequent chemical processing. The disk around the luminous Herbig star HD~179218 ($L_* \sim 100\,L_\odot$, $M_* \sim 3.0\,M_\odot$) offers a rare test case of CO-based mass measurements. Its CO snowline lies beyond the outer disk edge, and thus CO remains in the gas phase throughout. We present ALMA and NOEMA observations of six CO isotopologue lines ($^{13}$CO, C$^{18}$O, and C$^{17}$O, $J$=2--1 and 1--0). Fitting the spectral energy distribution yields $\mdust \approx 3 \times 10^{-5}\,\msun$. Modeling CO isotopologue emission with the thermochemical code DALI yields the gas-disk characteristic radius ($\rc \approx 90$~au) and the gas mass ($\mgas \approx 8 \times 10^{-3}\,\msun$). These results yield a gas-to-dust ratio of $\sim$270 for our fiducial dust opacity ($a_{\rm max} = 1$~mm), above the canonical interstellar value of $\sim$100. The thermal structure of our best-fitting model confirms that CO never freezes out in this disk, so the elevated ratio is not an artifact of CO depletion; together with the compact millimeter dust disk relative to the gas disk, it is consistent with substantial inward radial drift of millimeter dust. In addition, we find a gas cavity inside 50\,au that encloses the compact millimeter dust disk---the inverse of a typical transition disk---which we attribute to photoevaporative clearing of the inner gas after the pebbles drifted inward. HD~179218 thus provides a clean benchmark for studying the gas mass distribution and gas--dust interaction in protoplanetary disks, free from the CO-depletion biases that affect colder disks.
\end{abstract}

\section{Introduction}
\label{sec:introduction}

A fundamental question in planet formation is how much raw material is available in protoplanetary disks. The total disk mass---dominated by gas---determines the types of planets that can form and the formation timescale of gas giant planets. Yet measuring disk gas masses remains one of the most challenging problems in the field \citep{Bergin2017, Miotello2023}.

The primary constituent of disk gas, molecular hydrogen (H$_2$), lacks a permanent dipole moment and is essentially undetectable in the cold bulk of protoplanetary disks \citep[e.g.,][]{Bergin2017}. Gas masses are therefore most commonly inferred from rotational line emission of CO isotopologues. The rare isotopologues $^{13}$CO, C$^{18}$O, and C$^{17}$O are particularly useful: their low abundances keep them optically thin, so their emission probes the full gas column through the disk midplane \citep[e.g.,][]{Miotello2016, Miotello18, Williams2016, Zhang2017, Zhang2020}.

However, CO-based measurements face a serious systematic problem. Assuming an interstellar medium (ISM) CO abundance of $\sim$10$^{-4}$, ALMA surveys consistently find very low gas masses---often less than a Jupiter mass---and gas-to-dust ratios of $\sim$1--10 \citep[e.g.,][]{ansdell16, Miotello2016, Miotello18}. Independent constraints from HD emission \citep[e.g.,][]{Bergin2013, McClure2016} and stellar accretion rates \citep[][]{manara16} indicate that these CO-based masses underestimate the true gas content. Additional chemical tracers such as N$_2$H$^+$ confirm that CO is less abundant than expected even after accounting for freeze-out and photodissociation \citep[][]{Anderson19, Anderson22,  Trapman22, Trapman2025b}.

Interpreting CO as a gas-mass tracer is complicated by CO depletion, which we define as a reduction of the gas-phase CO abundance, relative to the interstellar value of $\sim$10$^{-4}$ per H$_2$, in the warm molecular layer, i.e., where $T_{\rm dust} \gtrsim 20$~K (above the CO freeze-out temperature) and below the CO photodissociation layer. Two mechanisms have been proposed in the literature to explain the lower-than-expected CO abundances in T~Tauri disks. One is chemical conversion, in which gas-phase CO is transformed into less volatile species such as CO$_2$, CH$_4$, and CH$_3$OH that remain frozen on grains even where CO itself would stay in the gas phase \citep{Schwarz2018,Bosman2018}. Another is dust evolution: CO freezes onto small grains in the upper disk, which then coagulate and settle, sequestering CO ice in large aggregates that are less accessible to observations \citep{Krijt2018,Krijt2020}. The evidence for substantial CO depletion is strongest for T~Tauri disks \citep[e.g.,][]{Miotello2017,Miotello2023}, whereas the warmer Herbig disks may be less affected \citep{Miotello2023,Trapman2025,Ricciardi2026}. So far, the magnitude and origin of the depletion remain debated \citep[e.g.,][]{Miotello2021,Ruaud2022,Powell2022,Deng2023,Zwicky2025}. Separately, photodissociation removes CO in the surface layers above the warm molecular layer. It is isotope-selective, destroying the rarer isotopologues preferentially because they self-shield less efficiently \citep{Miotello2016}, so it must be modeled explicitly when converting line emission to gas mass.

Crucially, both of these CO-depletion pathways require CO to freeze out first. In a disk where CO never freezes out, neither chemical conversion nor ice sequestration can operate, and the observable CO abundance should faithfully reflect the true gas content. This insight motivates our study of HD~179218.

HD~179218 is a Herbig Ae star with a luminosity of $\sim$100~$L_\odot$ (Table~\ref{tab:stellar_properties}). At this luminosity, we expect the CO snowline to lie well beyond the outer dust edge, such that CO should remain predominantly in the gas phase throughout the disk. This expectation is tested with thermochemical modeling in Section~\ref{sec:DALI_method}. If confirmed, HD~179218 provides an ideal system in which to measure the total gas mass free from the freeze-out-driven depletion that complicates measurements in cooler disks. The inner disk of HD 179218 has been studied extensively in the infrared: VLTI/MIDI interferometry revealed a pre-transitional structure with an inner disk on sub-au to few-au scales and an outer disk beyond $\sim$10~au \citep[][]{Fedele2008}, with subsequent VLTI/PIONIER, MIDI, and MATISSE modeling placing the inner/outer transition at $\sim$10~au \citep[][]{Kokoulina2021}. The disk is also notable for its strongly flared geometry and spatially extended polycyclic aromatic hydrocarbon emission \citep[][]{Taha2018}. However, none of these previous observations probed the bulk dust and gas reservoir of the outer disk. 

In this paper, we present NOEMA and ALMA observations of six CO isotopologue transitions in the HD~179218 disk. We first constrain the dust mass from millimeter continuum emission by fitting the spectral energy distribution (SED) (Section~\ref{sec:SED_model}). We then use the thermochemical code DALI \citep{bruderer2012,bruderer2013} to model the CO line emission, constraining the gas-disk characteristic radius and the gas mass jointly with a model grid fit to the optically thin C$^{18}$O and C$^{17}$O emission, and obtain a robust gas mass measurement for a disk in which CO depletion is negligible (Section~\ref{sec:DALI_method}). We discuss the implications of the resulting gas-to-dust ratio for dust evolution and the broader reliability of CO as a gas mass tracer in Section~\ref{sec:discussion}.

\begin{table}[h!]
      \centering
      \caption{Stellar Properties of HD~179218. Two published spectral types are listed. The other parameters are adopted from the compilation of \citet{Vioque2018}, which incorporates literature atmospheric parameters and Gaia DR2 astrometry.}
      \label{tab:stellar_properties}
      \small
      \begin{tabular}{l c l}
        \hline
        \hline
         Property & Value & Reference \\
         \hline
         Spectral Type & B9/A0 & \citep{Montesinos2009} \\
          &  & \citep{Vieira2003} \\
         $T_{\mathrm{eff}}$ (K) & 9500 & \citep{Vioque2018} \\
         Age (Myr) & $1.66^{+0.54}_{-0.26}$ & \citep{Vioque2018} \\
         $M_*/\msun$ & $3.0\pm0.2$ & \citep{Vioque2018} \\
         $\log(L_*/L_\odot)$ & $2.05^{+0.09}_{-0.14}$ & \citep{Vioque2018} \\
         Distance (pc) & 266 &  \citep{Gaia2018} \\
         \hline
      \end{tabular}
  \end{table}

\section{Observations}
\label{sec:observations}

\subsection{ALMA Observations}
\label{sec:alma_obs}

We observed HD~179218 in ALMA Band~3 (project 2021.1.0709.S; PI: Trapman) between 2021 October~24 and November~7, with baselines from 41\,m to 8.5\,km. Four spectral windows targeted the $^{13}$CO, C$^{18}$O, C$^{17}$O $J$=1--0, and CS $J$=2--1 lines at 110.201, 109.782, 112.359, and 97.981~GHz, respectively, each with a spectral resolution of $\sim$0.4~km~s$^{-1}$. A fifth spectral window centered at 100.5~GHz with a bandwidth of 1.875~GHz was used for continuum observations. The CS $J$=2--1 line is not used in the present analysis, which focuses on the CO isotopologues. Quasars J1914+1636 and J1924+1540 served as bandpass and phase calibrators. The total on-source integration time was 2.8~hours.

\begin{figure*}[t]
    \centering
    \includegraphics[width = 0.8\textwidth]{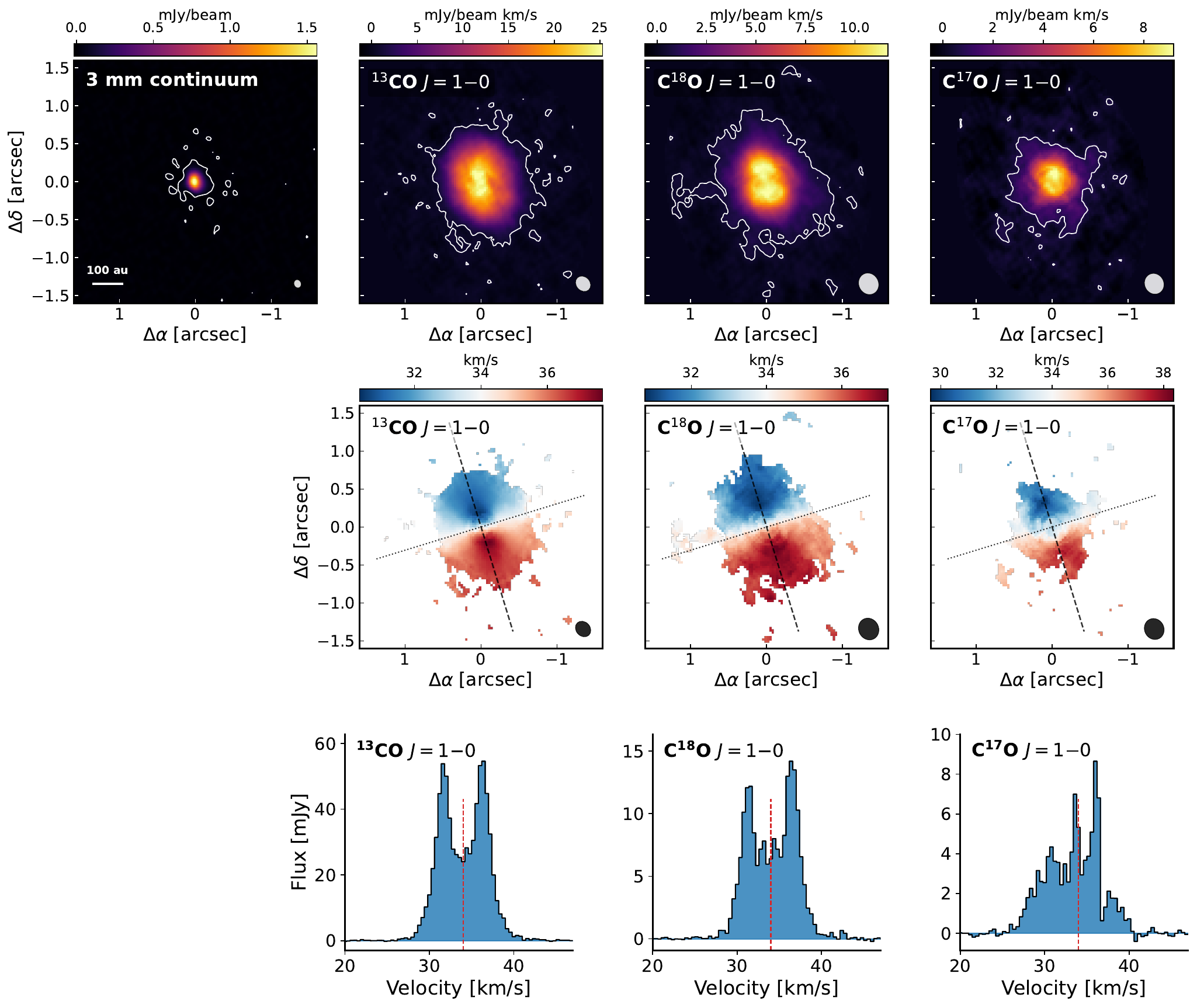}
    \vspace{-0.5em}
    \caption{ALMA Band 3 observations of HD 179218. Top row: Moment 0 (integrated intensity) maps of the 3 mm continuum (mJy/beam), $^{13}$CO $J=1{-}0$, C$^{18}$O $J=1{-}0$, and C$^{17}$O $J=1{-}0$ (mJy/beam km/s). The white contour marks the $3\sigma$ level in all panels. Middle row: Moment 1 (intensity-weighted velocity) maps for the three CO isotopologues, shown where the moment 0 SNR exceeds 3. The color scale is centered on the systemic velocity ($v_\mathrm{LSR} = 34$ km/s). Dashed and dotted lines indicate the disk major (PA $= 197°$) and minor axes, respectively. Bottom row: Integrated spectra extracted with a Keplerian mask. }
    \label{fig:alma_gallery}
    \vspace{1em}
    \includegraphics[width = 0.9\textwidth]{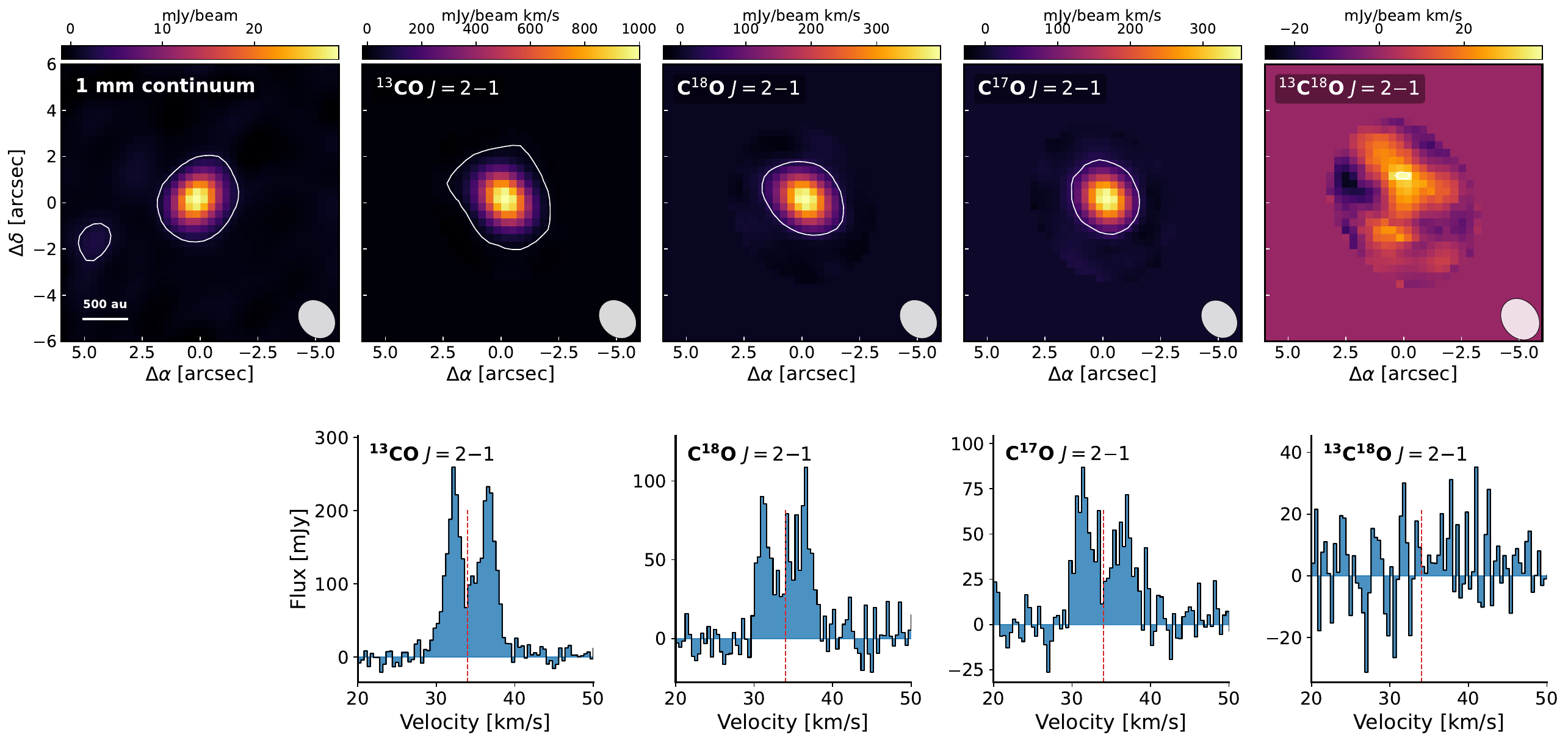}
    \vspace{-0.5em}
    \caption{NOEMA observations of HD 179218. Top row: Moment 0 (integrated intensity) maps of the 1.3 mm continuum (mJy/beam), $^{13}$CO, C$^{18}$O, C$^{17}$O, and $^{13}$C$^{18}$O $J=2{-}1$ (mJy/beam km/s). The white contour marks the $3\sigma$ level in all panels. Bottom row: Integrated spectra extracted within an elliptical mask with 2\arcsec\ radius. }
    \label{fig:noema_gallery}
\end{figure*}

The data were calibrated with the standard ALMA pipeline in CASA version 6.2.1 \citep{CASA2022}, followed by self-calibration using the procedure developed for the AGE-PRO large program \citep{Zhang_2025}. Images were produced with the \texttt{tclean} task using Briggs weighting (robust~$= 0.5$) and a channel width of 0.4~km~s$^{-1}$ for the line cubes, yielding synthesized beams of $\sim$0\farcs2 for the line data and $\sim$0\farcs1 for the continuum. Beam sizes, noise levels, and peak intensities for all images are listed in Table~\ref{tab:obs}. The resulting continuum and line maps are presented in Figure~\ref{fig:alma_gallery}.

\begin{deluxetable*}{lccccccc}
\tablecaption{Image parameters and integrated fluxes for HD\,179218. \label{tab:obs}}
\tablewidth{0pt}
\tablehead{
\colhead{Line} & \colhead{$\nu_{\rm rest}$} & \colhead{Beam} & \colhead{RMS} & \colhead{$S_{\rm peak}$} & \colhead{$S_{\rm int}$} & \colhead{$R_{68}$} & \colhead{$R_{90}$} \\
\colhead{} & \colhead{(GHz)} & \colhead{$B_{\rm maj}\times B_{\rm min}$ (B.P.A.)} & \colhead{} & \colhead{(mJy\,beam$^{-1}$)} & \colhead{(mJy\,km\,s$^{-1}$)} &
\colhead{(\arcsec)} & \colhead{(\arcsec)}
}
\startdata
\cutinhead{ALMA Band 3}
3\,mm cont & 97.7 & $0.11\times0.10$ ($21^\circ$) & 0.0083 & 1.646 & $4.38 \pm 0.25$\tablenotemark{a} & \nodata & \nodata \\
$^{13}$CO ($1{-}0$)       & 110.201 & $0.22\times0.18$ ($43^\circ$) & 0.239 & 11.16 & $303 \pm 3$ & $0.462 \pm 0.008$ & $0.653 \pm 0.019$\\
C$^{18}$O ($1{-}0$)       & 109.782 & $0.29\times0.26$ ($27^\circ$) & 0.143 & 5.08 & $78 \pm 2$ & $0.470 \pm 0.015$ & \nodata\\
C$^{17}$O ($1{-}0$)       & 112.359 & $0.28\times0.25$ ($27^\circ$) & 0.184 & 4.90 & $40 \pm 2$ & $0.421 \pm 0.035$ & \nodata \\
\cutinhead{NOEMA}
1.3\,mm cont  & 219.6 & $1.80\times1.47$ ($40^\circ$) & 0.498 & 30.73 & 31$\pm$1\tablenotemark{a} & \nodata & \nodata\\
$^{13}$CO ($2{-}1$)        & 220.399 & $1.82\times1.46$ ($41^\circ$) & 4.2 & 229.30 & $1209 \pm 22$ & \nodata & \nodata \\
C$^{18}$O ($2{-}1$)        & 219.560 & $1.81\times1.47$ ($41^\circ$) & 6.1 & 94.90 & $447 \pm 31$ & \nodata & \nodata \\
C$^{17}$O ($2{-}1$)        & 224.714 & $1.77\times1.43$ ($39^\circ$) & 6.7 & 78.40 & $402 \pm 35$ & \nodata & \nodata \\
$^{13}$C$^{18}$O ($2{-}1$) & 209.419 & $1.90\times1.55$ ($36^\circ$) & 6.2 & 40.63 & $40 \pm 31$ & \nodata & \nodata \\
\enddata
\tablecomments{
$R_{68}$ and $R_{90}$ are the deprojected radii enclosing 68\% and 90\% of the line flux, measured from the azimuthally averaged radial profiles. $R_{90}$ is reported only for $^{13}$CO, the line for which the curve of growth converges; for the fainter C$^{18}$O and C$^{17}$O lines $R_{90}$ is not well constrained.
}
\tablenotetext{a}{Integrated continuum flux density (mJy) within 2\arcsec radius.}
\end{deluxetable*}

\subsection{NOEMA Observations}
\label{sec:noema_obs}

We also observed HD~179218 at 1.3~mm with NOEMA (project S20AR; PI: Zhang) between 2020 June~11 and 17, using baselines from 32\,m to 148\,m. The correlator was configured to cover the $^{13}$CO, C$^{18}$O, C$^{17}$O, and $^{13}$C$^{18}$O $J$=2--1 lines at 220.399, 219.560, 224.714, and 209.419~GHz, with a native spectral resolution of $\sim$0.2~km~s$^{-1}$. Quasars 3C345 and J1932+204 were used for bandpass and phase calibration. The total on-source integration time was 3~hours.

Calibration was carried out with the standard NOEMA pipeline in GILDAS. Images were produced with the \texttt{MAPPING} task using natural weighting and a channel width of 0.4~km~s$^{-1}$. The resulting beam sizes and noise levels are listed in Table~\ref{tab:obs}, and the moment~0 maps and spectra are shown in Figure~\ref{fig:noema_gallery}.

\subsection{Integrated Line Fluxes}
\label{sec:line_fluxes}

Integrated line fluxes were measured using the curve-of-growth method \citep[][]{ansdell16, ansdell18, Andrews18, Sanchis2021}: the flux is summed within elliptical apertures of successively larger radius, matched to the disk inclination and position angle, until the enclosed flux converges to a stable plateau, which is adopted as the total flux. For the ALMA data, moment~0 maps were constructed with a Keplerian mask. The disk inclination and position angle (197$^\circ$) were determined from two-dimensional Gaussian fitting of the ALMA 3~mm continuum and $^{13}$CO $J$=1--0 moment~0 images, yielding an inclination of 38.7$^\circ$ that we round to 40$^\circ$ for the mask. The remaining mask parameters are a stellar mass of 3.0~$\msun$, distance of 266~pc, and systemic velocity of 34~km~s$^{-1}$.

For the NOEMA data, the disk is spatially unresolved, so we extracted fluxes within a fixed elliptical aperture with a 2\arcsec\ radius. Flux uncertainties were estimated from the per-beam rms in emission-free regions, scaled by the square root of the number of beams within the aperture. The measured fluxes are listed in Table~\ref{tab:obs}. The $^{13}$C$^{18}$O $J$=2--1 line is not detected and therefore is not used in the analysis.

\section{Dust Continuum and SED Modeling}
\label{sec:SED_model}
\label{sec:SED_results}

Our analysis proceeds in two stages: we first constrain the dust properties and dust mass from the spectral energy distribution (SED; this section), and then use them as inputs to thermochemical models of the CO line emission to measure the gas mass (Section~\ref{sec:DALI_method}).

Dust masses are commonly estimated directly from the millimeter flux density under the assumption of optically thin, isothermal emission at a single representative dust temperature. However, studies have shown that this approach may significantly underestimate the dust mass because part of the emission is optically thick \citep[e.g.,][]{Rilinger2023}. We instead determine $M_{\rm dust}$ from full radiative transfer modeling of the SED, which self-consistently computes the dust temperature structure. 

Wavelength-dependent dust opacities are computed using the \texttt{dsharp\_opac} code \citep[][]{Birnstiel2018}, which calculates absorption and scattering coefficients for a specified dust composition and grain size distribution. We adopt two dust populations, following \citet{Zhang2021}. The small grains ($a_{\rm max} = 1~\mu$m) are 50\% astronomical silicate \citep[][]{Draine2003} and 50\% refractory organics \citep[][]{Henning1996} by volume, with optical constants from the \texttt{dsharp\_opac} library, because water ice is expected to be photodesorbed in the UV-irradiated disk atmosphere. The large grains ($a_{\rm max} = 1$~mm or 1~cm) have the full DSHARP composition \citep[][]{Birnstiel2018}, including water ice and troilite. The large grains hold 95\% of the dust mass and the small grains 5\%, and $\mdust$ refers to the total dust mass. Grains in each population are compact, with $n(a) \propto a^{-3.5}$ \citep[][]{Mathis1977} from $a_{\rm min} = 0.005~\mu$m. Scattering is not included (\texttt{scattering\_mode\_max = 0} in \texttt{RADMC-3D}), so the opacity is purely absorptive. The millimeter opacity is set by the large grains, so we compute opacities for $a_{\rm max} = 1$~mm and 1~cm. The calculated opacities are used as input to \texttt{RADMC-3D} \citep[][]{Dullemond2012}, which solves the dust radiative transfer equation using the Monte Carlo method of \citet{Lucy1999} and \citet{Bjorkman2001} to compute the dust temperature structure and synthetic SEDs. The star is represented by a Kurucz model atmosphere with $T_{\rm eff} = 9500$~K, scaled to the stellar luminosity in Table~\ref{tab:stellar_properties}. The disk density structure is parameterized in cylindrical coordinates with a vertical scale height
\begin{equation}
H(r) = H_{100} \left( \frac{r}{100\,\mathrm{au}} \right)^{\beta},
\end{equation}
where $H_{100}$ is the scale height at 100~au and $\beta$ is the flaring exponent of the scale height itself (so that the aspect ratio $H/r \propto r^{\beta-1}$; a flared disk therefore corresponds to $\beta > 1$). The three parameters varied in the SED fitting are $\beta$, $H_{100}$, and the dust mass, as these most strongly influence the continuum emission.

The observed SED is assembled from archival photometry retrieved through the VizieR photometry service, spanning the optical (Hipparcos Tycho, Gaia EDR3, and Johnson/Cousins $UBVRI$), near-infrared (2MASS and Johnson $JHKLM$), and mid- to far-infrared (WISE W1--W4, AKARI IRC and FIS, and IRAS 12--100~$\mu$m) bands, supplemented by our NOEMA 1.3~mm continuum flux (Table~\ref{tab:obs}). The dust mass is determined by fitting the resulting model SEDs to this photometry.

\begin{figure}[t]
    \centering
    \includegraphics[width=3.5in]{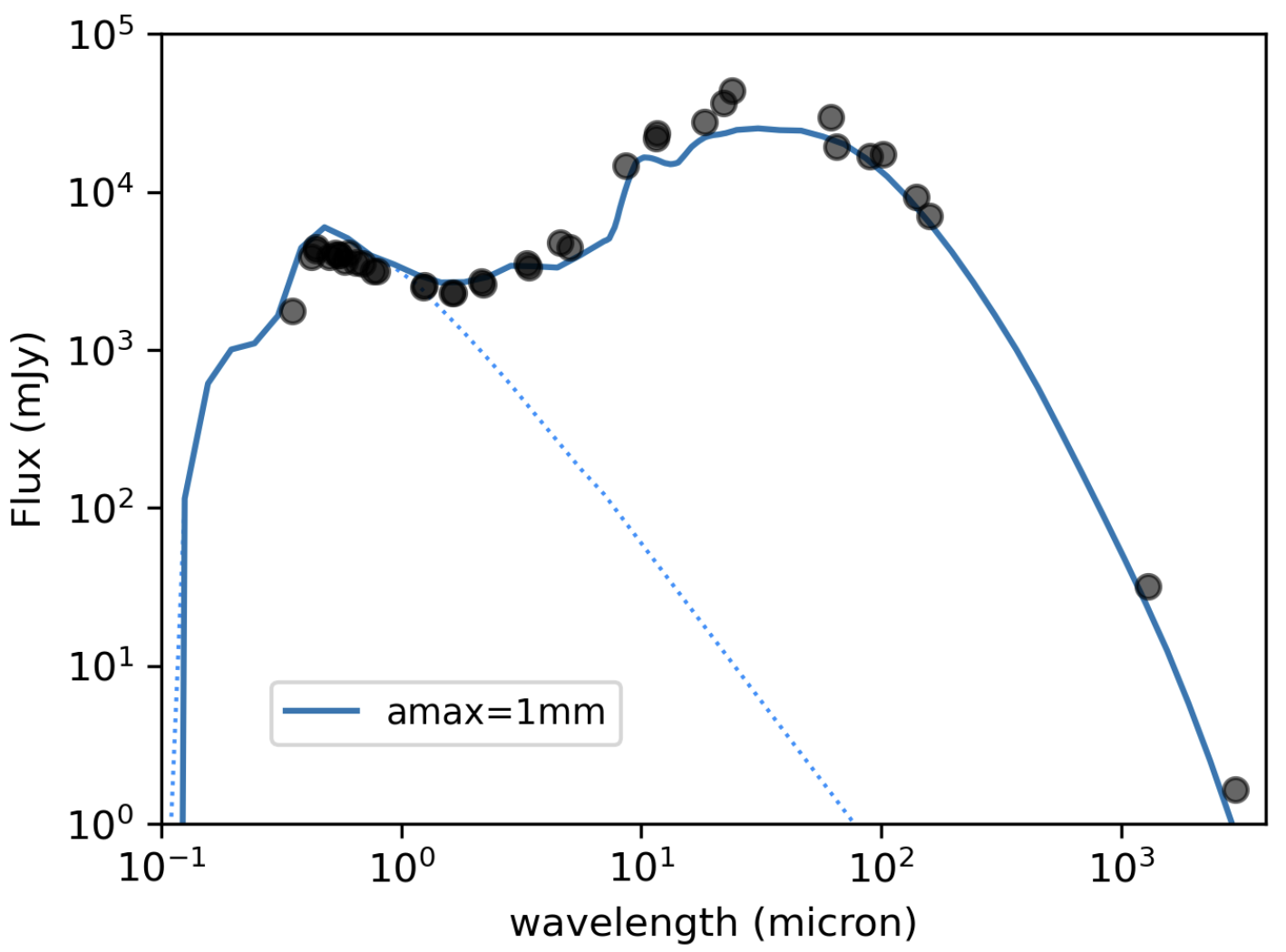}
    \caption{Observed SED of HD~179218 (black points) compared to the best-fitting \texttt{RADMC-3D} model assuming a maximum grain size of 1~mm. The dashed blue line shows the stellar photosphere model.}
    \label{fig:SED_fit}
\end{figure}

The models with the two adopted maximum grain sizes fit the SED comparably well and favor $\beta = 1.10$ and $H_{100} = 10$\,au. We adopt $a_{\rm max} = 1$~mm as fiducial (Figure~\ref{fig:SED_fit}) because it is commonly used, which eases comparison with the literature. This gives $\mdust = 3.0 \times 10^{-5}\,\msun$, while $a_{\rm max} = 1$~cm requires $6.0 \times 10^{-5}\,\msun$. The gas mass is set by the CO isotopologue emission and, to first order, does not depend on the dust opacity, so a larger dust mass lowers the gas-to-dust ratio rather than raising $\mgas$ (Section~\ref{sec:radial_drift}).

The dust mass also depends on the grain-size distribution, composition, and other dust properties, which can change it by factors of a few \citep[e.g.,][]{Liu2022, Xin2023, Liu2024} and are not captured by the formal flux-calibration uncertainty. The fiducial model also lies slightly below the observed emission at $\gtrsim 100~\mu$m (Figure~\ref{fig:SED_fit}), so a better fit there could raise $\mdust$. We therefore treat $\mdust = 3.0 \times 10^{-5}\,\msun$ as a best-fit value under our dust model. Together with $H_{100}$ and $\beta$, it is a fixed input to the gas modeling (Table~\ref{tab:disk_properties}).

\section{Thermochemical Modeling with DALI}
\label{sec:DALI_method}

We use the thermochemical code DALI \citep[][]{bruderer2012, bruderer2013} to model the CO isotopologue line emission from the HD~179218 disk. DALI self-consistently calculates the thermal and chemical structure of a 2D axisymmetric disk given the stellar spectrum and disk density structure. The code first solves the dust radiative transfer to obtain the dust temperature and UV radiation field, then iteratively solves for the gas temperature, chemistry, and molecular excitation until thermal balance is reached. The chemistry is evolved for 1~Myr before the molecular abundances are extracted; in this warm disk, where CO never freezes out, the gas-phase CO abundance reaches steady state well within this time, so the derived abundances are insensitive to the exact chemical age. The converged model is ray-traced at a distance of 266~pc and an inclination of 38.7$^\circ$ (Section~\ref{sec:line_fluxes}) to produce synthetic spectral image cubes.

Crucially for this work, DALI treats each CO isotopologue with its own chemical network rather than scaling a single CO abundance by fixed elemental ratios \citep[][]{Miotello2014, Miotello2017}. Because the rarer isotopologues $^{13}$CO, C$^{18}$O, and C$^{17}$O are less abundant, they self-shield less efficiently than $^{12}$CO and are photodissociated deeper into the disk. As a result, their abundances deviate from constant ratios relative to $^{12}$CO, and their emitting regions differ. This isotope-selective treatment is essential for using the optically thin C$^{18}$O and C$^{17}$O lines as quantitative gas-mass tracers.

DALI adopts a viscous accretion disk surface density profile \citep[][]{Lynden1974, Hartmann1998}:
\begin{equation}
      \Sigma_{\rm gas} = \Sigma_{c} \left(\frac{R}{\rc}\right)^{-\gamma} \exp\left[-\left(\frac{R}{\rc}\right)^{2-\gamma}\right],
\end{equation}
where $\rc$ is the gas-disk characteristic radius and $\Sigma_c$ is related to the total gas mass by
\begin{equation}
      \Sigma_{c} = (2-\gamma)\frac{\mgas}{2\pi {\rc}^2}.
\end{equation}
The dust surface density is set by the gas-to-dust mass ratio. The model extends from the sublimation radius, $R_{\rm subl} \sim 0.07\sqrt{L_*/L_\odot}$~au \citep[][]{Dullemond2001}, interior to which both gas and dust densities are set to zero, out to an outer radius of 300~au.

\subsection{Model Grid}

We search for the best-fitting model by running a grid over the disk parameters. The stellar parameters and the disk vertical structure ($H_{100}$, $\beta$) are fixed at the values from the SED fit (Section~\ref{sec:SED_results}). We then vary the gas-disk characteristic radius $\rc$ and the gas-to-dust mass ratio (and hence the gas mass) as the primary grid parameters; the full grid is summarized in Table~\ref{tab:dali_grid}. The {gas-disk characteristic radius primarily sets the spatial extent of the line emission, while the gas-to-dust mass ratio sets the column density and thus the flux of the optically thin C$^{18}$O and C$^{17}$O lines.

Our first round of model exploration shows that a single full-disk profile that fits the outer disk always significantly overpredicts the line emission interior to $\sim$50~au, so we introduce a central cavity to reduce the surface density in the inner disk. We tested cavity radii of 15, 30, and 50~au; a 50~au cavity provides the best match to the resolved $J$=1--0 profiles. The depth of the cavity is set by a depletion factor $\delta$, defined as the factor by which the gas surface density drops at the outer edge of the cavity relative to the extrapolation of the outer-disk profile: the surface density interior to the cavity radius is multiplied by $\delta$, so $\delta = 1$ corresponds to no cavity and smaller $\delta$ to a deeper cavity. We additionally explored the surface density power-law index $\gamma$. The effect of each parameter on the line emission is summarized in Table~\ref{tab:dali_grid}.

\begin{table*}[t]
\centering
\caption{DALI model grid for the gas modeling. Fixed parameters are held at the listed value; grid parameters are stepped through the listed values. The depletion factor $\delta$ is the factor by which the gas surface density drops at the outer edge of the cavity and applies only to models that include a cavity.\label{tab:dali_grid}}
\begin{tabular}{l l l}
\hline
\hline
Parameter & Value(s) & Type \\
\hline
$M_*$, $L_*$, distance        & Table~\ref{tab:stellar_properties} & fixed (stellar) \\
$H_{100}$, $\beta$            & SED fit (Section~\ref{sec:SED_results}) & fixed (SED) \\
$\mdust$                      & SED fit (Section~\ref{sec:SED_results}) & fixed (SED) \\
\hline
$\rc$ (au)                    & 15, 20, 30, 45, 60, 90, 120, 150 & grid \\
gas-to-dust ratio            & 10, 30, 100, 150, 240, 270, 300, 450, 1000 & grid \\
\hline
cavity radius (au)           & none, 15, 30, 50 & explored \\
depletion factor $\delta$    & 0.01, 0.03, 0.1, 0.15, 0.2, 0.3 & explored \\
$\gamma$                     & 0.5, 1.0 & explored \\
\hline
\end{tabular}
\end{table*}

\subsection{Comparison to Observations}

Because our goal is to measure the gas mass, we base the goodness of fit on the observables that trace the gas column density: the optically thin C$^{18}$O and C$^{17}$O lines. The optically thick $^{13}$CO emission instead probes the gas temperature in the disk surface layers rather than the column density, so it carries little information about the mass; we use it as a consistency check on the models rather than including it in the fit.

For the ALMA $J$=1--0 data, we compare azimuthally averaged radial intensity profiles between the models and observations, using the C$^{18}$O $J$=1--0 profile. The C$^{17}$O $J$=1--0 profile is not used because the transition has three hyperfine components whose blending, combined with the limited resolution of the images in the inner region, contaminates the radial profile extracted from the moment~0 map (Appendix~\ref{app:excluded_lines}, Figure~\ref{fig:C17O_best_radial}). For the NOEMA $J$=2--1 data, we compare the integrated line spectra for C$^{18}$O and C$^{17}$O; at this higher frequency the hyperfine splitting of C$^{17}$O is small relative to the line width, so the line can be included. The goodness of fit is quantified using $\chi^2$, computed as the sum over these three measurements (the C$^{18}$O $J$=1--0 radial profile and the C$^{18}$O and C$^{17}$O $J$=2--1 spectra).

The $^{13}$CO comparison yields a noteworthy result: across the full parameter coverage of our grid, models that reproduce the C$^{18}$O emission always overpredict the $^{13}$CO $J$=1--0 and $J$=2--1 fluxes. As we show in Section~\ref{sec:disk_size}, this reflects an overestimate of the gas temperature in the $^{13}$CO-emitting surface layer rather than an error in the disk size or mass, and the $^{13}$CO radial profile shape remains a useful consistency check on the disk size.

\subsection{Constraining the Disk Size}
\label{sec:disk_size}

The disk size constraint comes mainly from the C$^{18}$O $J$=1--0 radial profile, with the C$^{18}$O and C$^{17}$O $J$=2--1 spectra providing additional constraints.

\begin{figure*}[htbp]
\centering
\includegraphics[width=0.8\textwidth]{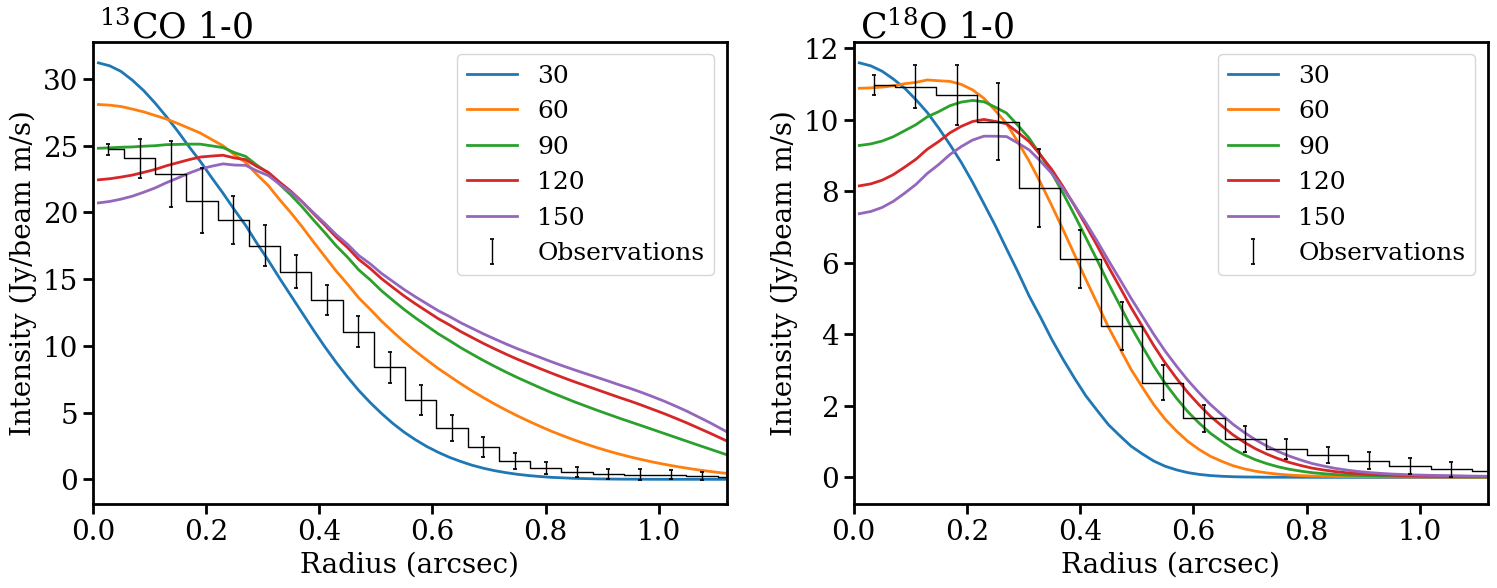}
\caption{Effect of varying the gas-disk characteristic radius on the $^{13}$CO and C$^{18}$O $J$=1--0 radial profiles. ALMA observations are shown in black. The optically thin C$^{18}$O profile is best matched by $\rc = 90$~au: smaller radii underpredict the spatial extent of the emission and larger radii are too extended. The optically thick $^{13}$CO flux is overpredicted by all models (Section~\ref{sec:DALI_method}) and constrains only the spatial extent, which is consistent with $\rc = 90$~au. All models include a 50~au cavity except the $\rc = 30$~au profile, which has no cavity because the cavity radius would exceed the gas-disk characteristic radius.}
\label{fig:Rc_comparison}
\end{figure*}

Figure~\ref{fig:Rc_comparison} shows the comparison between observed and modeled radial profiles for different values of $\rc$. The C$^{18}$O $J$=1--0 profile, which is optically thin and traces the mass-bearing gas, is best reproduced by $\rc \approx 90$~au: models with $\rc < 60$~au underpredict the spatial extent of the emission, while models with $\rc \gtrsim 100$~au are more extended at outer radii, while underpredicting inner radii. The NOEMA C$^{18}$O and C$^{17}$O $J$=2--1 integrated fluxes provide a consistent constraint, with models of $\rc < 60$~au underpredicting them. The optically thick $^{13}$CO emission is excluded from the gas-mass fit because it is mostly sensitive to the gas temperature in the atmosphere instead of the bulk gas mass (Section~\ref{sec:DALI_method}). The $^{13}$CO $J$=1--0 flux is overpredicted by all models that reproduce the C$^{18}$O $J$=1--0 and is therefore excluded from the fit, although its spatial extent is consistent with $\rc \approx 90$~au. We adopt $\rc = 90$~au for the gas mass determination.

The nature of the $^{13}$CO overprediction points to its cause. The model reproduces the \textit{shape} of the observed radial profile but lies above it in intensity at all radii, and the same offset is seen in the $J$=2--1 spectrum (Appendix~\ref{app:excluded_lines}, Figure~\ref{fig:13CO_best_spectra}). Because the $^{13}$CO lines are optically thick, their surface brightness traces the gas temperature at the $\tau = 1$ surface (Figure~\ref{fig:tgas_co}); a uniform overprediction with the correct shape therefore indicates that the model gas temperature in this layer is too high. This may reflect overestimated photoelectric and PAH heating in the intense UV field of this luminous Herbig star, or an inaccurate vertical structure, including the scale height and emitting-layer location \citep{Ruaud2022,PanequeCarreno2025}. In contrast, the optically thin C$^{18}$O and C$^{17}$O lines probe the mass-bearing gas column rather than the surface temperature, so the $^{13}$CO overprediction does not affect the gas mass. This decoupling is not unique to HD~179218: even faint, low-mass disks can have optically thick $^{12}$CO and $^{13}$CO \citep{Ricciardi2026}.

\subsection{Measuring the Gas Mass}
\label{sec:gas_mass}

Within the grid, the gas-to-dust mass ratio (and hence the gas mass) controls the flux of the optically thin C$^{18}$O and C$^{17}$O lines, whose emission scales directly with the gas column density. We constrain it from the C$^{18}$O $J$=1--0 radial profile together with the C$^{18}$O and C$^{17}$O $J$=2--1 spectra, the same optically thin tracers that enter our $\chi^2$ (Section~\ref{sec:DALI_method}).

\begin{figure*}[htbp]
\centering
\includegraphics[width=0.8\textwidth]{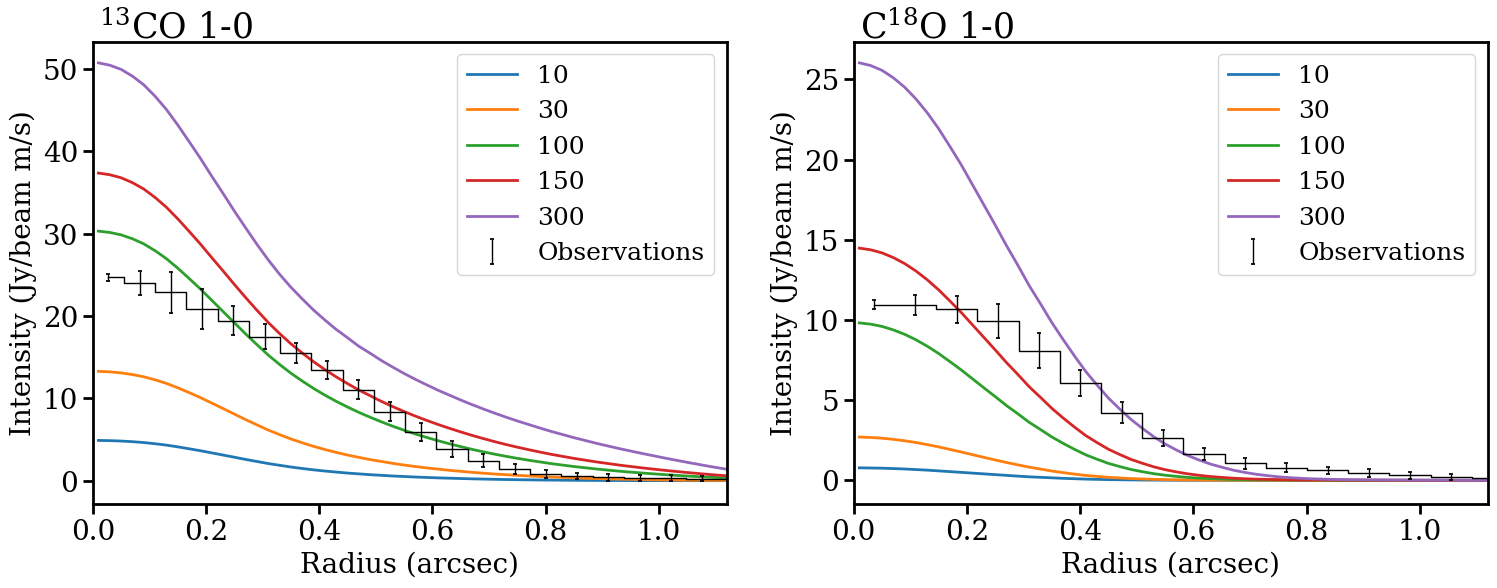}
\caption{Effect of varying the gas-to-dust mass ratio on the C$^{18}$O and $^{13}$CO $J$=1--0 radial profiles, for a fixed $\rc = 90$~au and no cavity. The models shown isolate the effect of the gas-to-dust ratio while keeping the other model parameters fixed. The gas-to-dust ratio of 270 is not shown because the best-fitting model with this ratio includes a 50~au cavity; it is therefore not directly comparable to the no-cavity models shown here. The C$^{18}$O emission, which sets the gas mass, is best matched by a gas-to-dust ratio of $\sim$270. The $^{13}$CO profile is shown for context but is overpredicted by the models that best reproduce the C$^{18}$O emission, whereas models with lower gas-to-dust ratios (10 and 30) underpredict the inner $^{13}$CO emission; optically thick $^{13}$CO is excluded from the fit (Section~\ref{sec:DALI_method}). The residual overprediction of both lines at small radii motivates the inclusion of a cavity (Figure~\ref{fig:gap_depletion}); the best-fit gas-to-dust ratio remains $\sim$270 when the cavity is included (Section~\ref{sec:best_fit}).}
\label{fig:gtd_comparison}
\end{figure*}

Figure~\ref{fig:gtd_comparison} compares models with different gas-to-dust mass ratios to the observed radial profiles. A gas-to-dust ratio of $\sim$270 provides the best match to the optically thin C$^{18}$O emission, and the same value reproduces the C$^{18}$O and C$^{17}$O $J$=2--1 spectra. Lower ratios ($\lesssim$100) significantly underpredict the line flux, while higher ratios ($\gtrsim$1000) overpredict it. Combined with the fiducial dust mass of $3.0 \times 10^{-5}\,\msun$, this yields a gas mass of
$\mgas \approx 8.1 \times 10^{-3}$\,\msun.

A parabolic fit to $\chi^2$ across the three finest grid points (gas-to-dust = 240, 270, and 300), which bracket the minimum, yields a formal fitting uncertainty of only $\pm 5$ ($1\sigma$, $\Delta\chi^2 = 1$). The error budget is instead dominated by the $\sim$10\% absolute flux calibration, which is correlated across channels and propagates linearly to the optically thin line fluxes, giving a gas-to-dust ratio of $270 \pm 28$ and $\mgas = (8.1 \pm 0.8) \times 10^{-3}\,\msun$ at fixed dust mass. The gas mass is set by the CO isotopologue emission and is, to first order, independent of the dust opacity. The dominant systematic on the gas-to-dust ratio is instead the dust opacity: a maximum grain size of 1~cm instead of 1~mm doubles the dust mass (Section~\ref{sec:SED_results}) and hence halves the gas-to-dust ratio to $\sim$135, which is still above the ISM value of $\sim$100.

\subsection{Best-Fitting Model}
\label{sec:best_fit}

\begin{figure*}[htbp]
\centering
\includegraphics[width=0.8\textwidth]{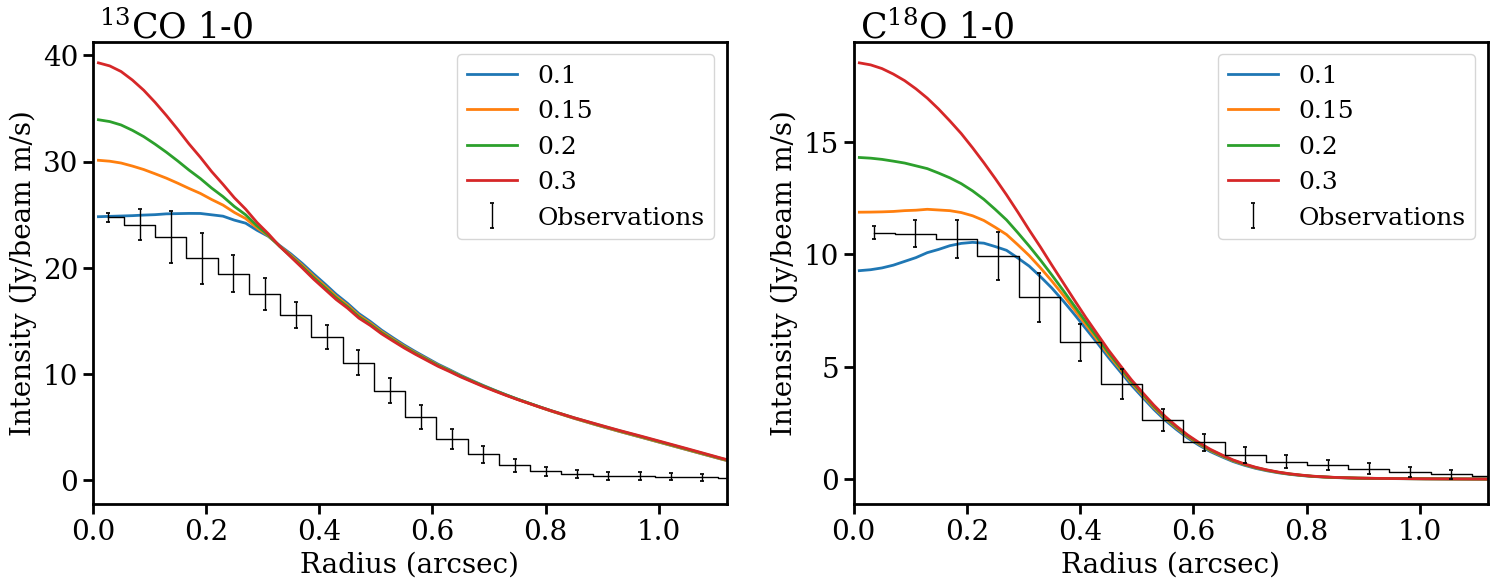}
\caption{Effect of the 50~au cavity and the surface density depletion factor $\delta$ on the $^{13}$CO and C$^{18}$O $J$=1--0 radial profiles, compared to the ALMA observations (black). Models without a cavity overpredict the emission in the innermost region; introducing a cavity at 50~au with $\delta = 0.15$ suppresses the inner emission and reproduces the observed profiles.}
\label{fig:gap_depletion}
\end{figure*}

The cavity at 50~au is required by the resolved C$^{18}$O $J$=1--0 radial profile, with the $^{13}$CO profile showing the same inner-region behavior. As shown in Figure~\ref{fig:gap_depletion}, models without a cavity systematically overpredict the line emission in the innermost region, whereas introducing a cavity at 50~au with a gas surface density depletion factor $\delta = 0.15$ suppresses the inner emission and brings the model profiles into agreement with the observations. This cavity is well outside the $\sim$10~au inner-disk structure inferred from infrared interferometry \citep[][]{Fedele2008, Kokoulina2021} and is resolved directly by our ALMA observations.

\begin{figure*}[t]
    \centering
    \includegraphics[width=\textwidth]{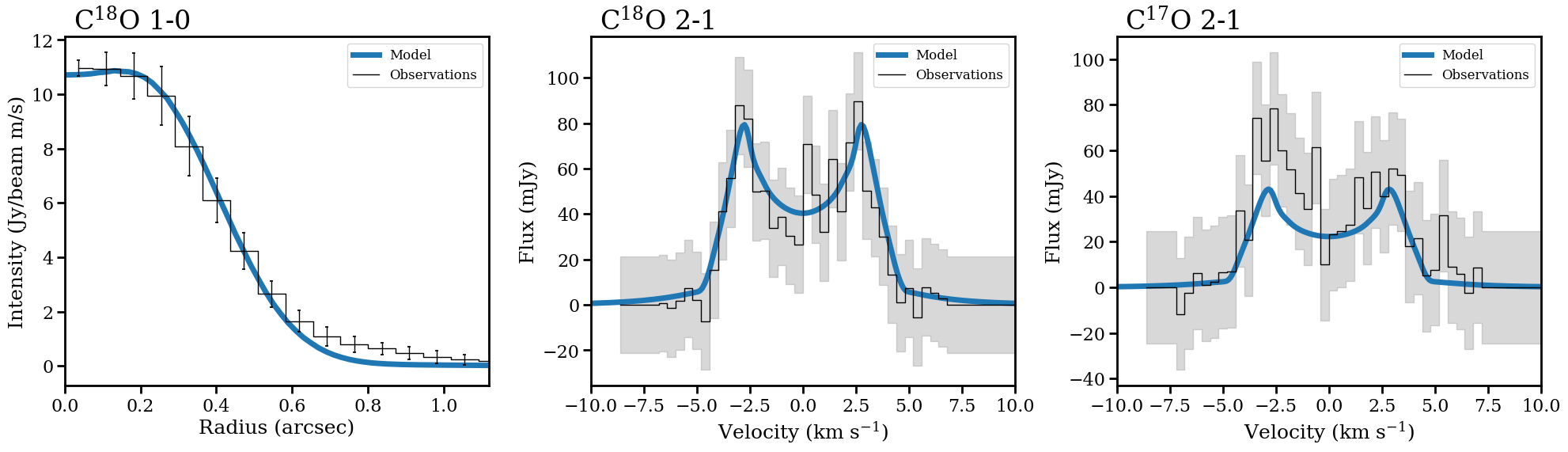}
    \caption{Best-fitting DALI model (blue) compared to the three observed measurements included in the $\chi^2$ (black): the ALMA C$^{18}$O $J$=1--0 radial profile and the NOEMA C$^{18}$O and C$^{17}$O $J$=2--1 spectra. The model has $\rc = 90$~au, a gas-to-dust mass ratio of 270, a flaring index $\beta = 1.1$, and a cavity at 50~au with a surface density depletion factor $\delta = 0.15$.}
    \label{fig:best_models}
\end{figure*}

With the cavity included, the best-fitting DALI model is characterized by an outer-disk gas-to-dust mass ratio of 270 ($\mgas = 8.1 \times 10^{-3}\,\msun$), a gas-disk characteristic radius of 90~au, a cavity at 50~au with a surface density depletion factor $\delta = 0.15$, the flaring index fixed from the SED fit ($\beta = 1.1$, i.e. $H \propto r^{1.1}$), and a surface density power-law index $\gamma = 1.0$, which is preferred over $\gamma = 0.5$ by the $\chi^2$ comparison. The total $\chi^2$, computed from the three optically thin tracers that constrain the gas mass, is 49.5, with individual contributions of 23.6 (C$^{18}$O $J$=1--0 radial profile), 9.8 (C$^{18}$O $J$=2--1 spectrum), and 16.2 (C$^{17}$O $J$=2--1 spectrum). This model gives the lowest $\chi^2$ in the grid. As noted in Section~\ref{sec:DALI_method}, the optically thick $^{13}$CO $J=1$–0 and $J=2$–1 lines are excluded from the $\chi^2$ calculation because their emission is less directly sensitive to the total gas mass than the optically thin C$^{18}$O emission.

Figure~\ref{fig:best_models} shows the comparison between the best-fitting model and the three measurements included in the $\chi^2$: the ALMA C$^{18}$O $J$=1--0 radial profile and the NOEMA C$^{18}$O and C$^{17}$O $J$=2--1 spectra. The disk properties constrained in this work are summarized in Table~\ref{tab:disk_properties}.

\begin{table}[ht]
\centering
\caption{Disk Properties of HD~179218 Constrained in This Work. The quoted uncertainties on the dust mass, gas-to-dust ratio, and gas mass reflect the formal uncertainty associated with the $\sim$10\% continuum flux calibration (Section~\ref{sec:gas_mass}) and do not include systematic uncertainties associated with the adopted dust opacity, including the grain-size distribution, maximum grain size, composition, porosity, or uncertainties in the SED fit. The gas mass is given at fixed $\mdust$.} The dominant systematic on the gas-to-dust ratio is the dust opacity: $a_{\rm max} = 1$~cm instead of 1~mm doubles $\mdust$ and halves the gas-to-dust ratio to $\sim$135, while $\mgas$, which is constrained by the CO lines, is largely unchanged.\label{tab:disk_properties}
\begin{tabular}{l c}
\hline
\hline
Property & Value \\
\hline
Inclination & $38.7^\circ$ \\
Position angle & $197^\circ$ \\
Scale height $H_{100}$ & 10~au \\
Flaring index $\beta$ & 1.10 \\
$\mdust$ ($a_{\rm max} = 1$~mm) & $(3.0 \pm 0.3) \times 10^{-5}\,\msun$ \\
Gas-disk characteristic radius $\rc$ & 90~au \\
Surface density index $\gamma$ & 1.0 \\
Cavity radius & 50~au \\
Cavity depletion factor $\delta$ & 0.15 \\
Gas-to-dust ratio & $270 \pm 28$ \\
$\mgas$ & $(8.1 \pm 0.8) \times 10^{-3}\,\msun$ \\
\hline
\end{tabular}
\end{table}

\section{Discussion}
\label{sec:discussion}

\subsection{Negligible CO Freeze-out in the HD~179218 Disk}
\label{sec:no_freezeout}

The central premise of this work is that the high stellar luminosity of HD~179218 prevents CO from freezing out anywhere in the disk, removing the dominant source of systematic uncertainty in CO-based gas mass measurements. Our best-fitting DALI model confirms this directly (Figure~\ref{fig:tgas_co}). CO freeze-out is set by the dust temperature, but gas and dust are thermally coupled near the disk midplane, so the gas temperature shown here is a good proxy for $T_{\rm dust}$ there. The midplane gas temperature remains above the CO freeze-out temperature ($\sim$20~K) at all radii: it decreases from several hundred~K in the inner disk to $\sim$30~K at the outer edge of the model (300~au), passing through $\sim$50~K at 50~au and $\sim$40~K at the gas-disk characteristic radius of 90~au. As a result, CO retains its full gas-phase abundance ($n_{\rm CO}/n_{\rm gas} \sim 10^{-4}$) down to the midplane throughout the disk, in contrast to colder disks where the bulk of the CO column is frozen out in the outer midplane.

\begin{figure*}[t]
    \centering
    \includegraphics[width=\textwidth]{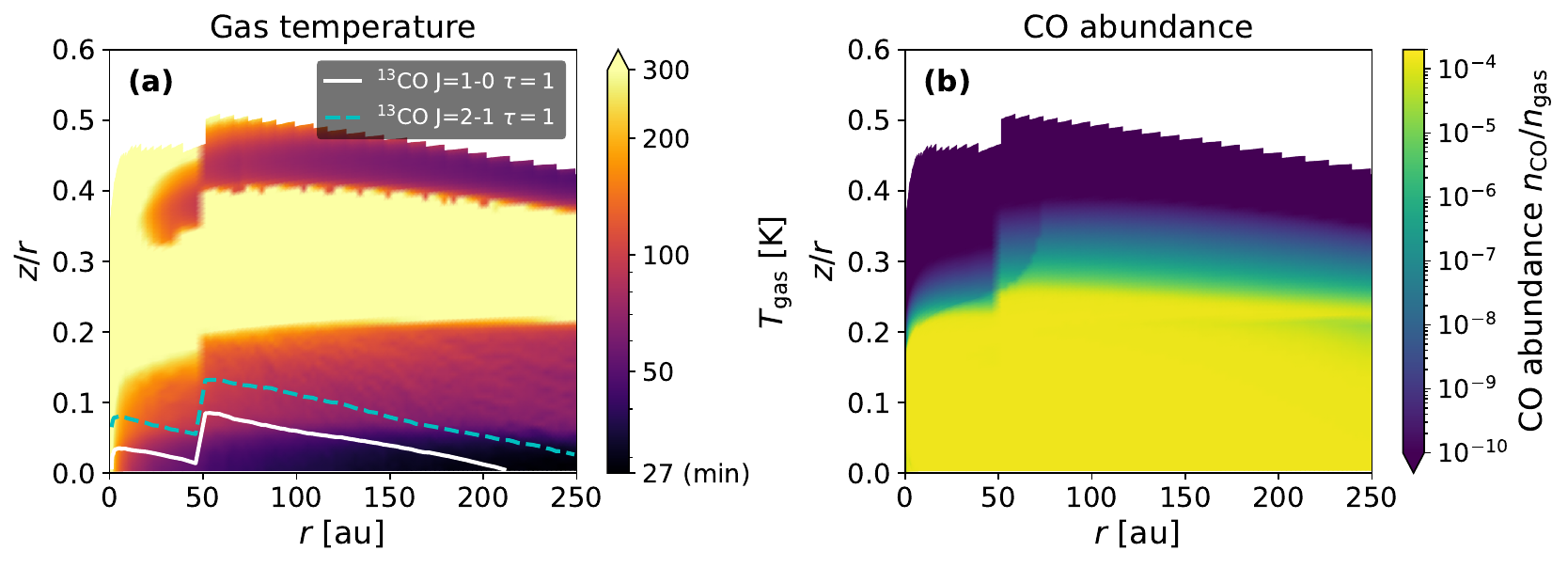}
    \caption{Two-dimensional structure of the best-fitting DALI model ($\rc = 90$~au, gas-to-dust ratio 270, cavity at 50~au). \textit{Left:} gas temperature (logarithmic color scale).  The color scale saturates at the minimum modeled gas temperature in the displayed region (27~K), which lies above the CO freeze-out temperature of $\sim$20~K. The solid white and dashed cyan curves mark the approximate $\tau = 1$ surfaces of the $^{13}$CO $J$=1--0 and $J$=2--1 lines, computed from the line-center optical depth along the line of sight using the level populations of the model; the optically thick $^{13}$CO emission arises from an elevated layer with $T_{\rm gas} \approx 30$--110~K. \textit{Right:} CO abundance ($n_{\rm CO}/n_{\rm gas}$). The midplane gas temperature remains above the CO freeze-out temperature ($\sim$20~K) at all radii, so CO stays in the gas phase at its full abundance ($\sim$10$^{-4}$) down to the midplane throughout the disk.}
    \label{fig:tgas_co}
\end{figure*}

In this work, we adopt ``CO depletion" as defined in Section~\ref{sec:introduction}: a reduction of the gas-phase CO abundance in the warm molecular layer ($T_{\rm dust} \gtrsim 20$~K, below the photodissociation layer), arising from freeze-out-related processes, namely conversion of CO to less volatile species and sequestration of CO ice onto settled grains. The $\sim$20~K threshold is approximate, and the photodissociation boundary lies deeper for the rarer isotopologues. 


For HD~179218, the DALI thermal structure shows no CO freeze-out. Therefore, neither pathway can reduce the bulk gas-phase CO abundance, and our gas mass is not subject to the CO depletion that biases estimates in cooler disks. Isotope-selective photodissociation, a separate process in the surface layers, is still present and is included self-consistently in DALI through isotopologue-specific chemistry and self-shielding.

\subsection{Evidence for Dust Evolution and Radial Drift}
\label{sec:radial_drift}

In the standard picture of dust evolution, particles grow to millimeter and centimeter sizes in the outer disk, decouple from the gas, and drift inward due to aerodynamic drag, selectively removing solid material from the outer disk while leaving the gas largely unaffected \citep[e.g.,][]{Trapman2020}. Three lines of evidence, each with different systematics, are consistent with substantial inward drift of millimeter dust in the HD~179218 disk.

First, the measured gas-to-dust mass ratio of $270 \pm 28$ exceeds the ISM value of $\sim$100. Because CO freeze-out is absent, the gas mass should not be underestimated by CO depletion, so the elevated gas-to-dust ratio is consistent with a relative deficit of millimeter dust. Its interpretation depends on the dust opacity: for $a_{\rm max} = 1$~cm, the dust mass doubles and the ratio drops to $\sim$135. We therefore take the ratio as consistent with substantial dust evolution, not as independent evidence for radial drift.

Second, the gas disk is substantially more extended than the millimeter dust disk. A two-dimensional Gaussian fit to the ALMA 3~mm continuum yields a beam-deconvolved size of $\sim$0\farcs14, corresponding to a dust emission radius of only $\sim$20--40~au, whereas the $^{13}$CO $J$=1--0 emission has $R_{68} = 0\farcs46$ ($\sim$120~au; Table~\ref{tab:obs}). The resulting gas-to-dust size ratio of $\gtrsim$3 is consistent with substantial radial drift \citep[e.g.,][]{Trapman2020}, and is insensitive to the dust opacity assumption. Because $^{13}$CO is more optically thin than $^{12}$CO and therefore traces a smaller radius, this ratio is a conservative lower limit on the value that would be obtained from the $^{12}$CO emission commonly used in the literature.

A spatially resolved comparison of the continuum and gas radial profiles would characterize the difference in spatial extent more directly than a single Gaussian size. It would show how strongly millimeter dust is depleted from the outer disk and constrain how much solid material has moved inward \citep[e.g.,][]{Testi2014,Deng2025}.

However, the size difference alone cannot uniquely determine the amount of radial drift, since fragmentation can also replenish small, gas-coupled grains and reshape the continuum profile \citep[e.g.,][]{Testi2014,Deng2025}. The compact millimeter dust emission relative to the gas indicates substantial dust evolution and is consistent with radial drift, but quantifying the inward transport would require resolved continuum profiles and models of growth, fragmentation, and drift.

Third, the continuum spectral index between the ALMA 3~mm and NOEMA 1.3~mm integrated fluxes is $\alpha \approx 2.5$, consistent with the presence of grains that have grown to at least millimeter sizes and subsequently drifted inward, rather than a population of small, ISM-like grains. Taken together, these three indicators are consistent with substantial radial drift of the millimeter dust in this disk.

\subsection{The Origin of the Inner Gas Depletion}
\label{sec:cavity_origin}

The disk structure inferred from our modeling is unusual: the millimeter dust emission is compact ($\sim$20--40~au) and sits \textit{inside} the 50~au region where the gas surface density is depleted ($\delta = 0.15$), and no millimeter ring is detected at the cavity edge despite our $\sim$0\farcs1 resolution (50~au corresponds to 0\farcs19 at 266~pc). This is the opposite of the configuration seen in typical transition disks, where millimeter dust accumulates in a ring at the pressure maximum just outside the gas cavity.

We suggest that this structure records a two-step evolution: rapid radial drift followed by photoevaporation-driven clearing of the inner gas. Radial drift operates from early times and is fast in this disk: for our best-fit surface density, millimeter grains at $\sim$100~au have Stokes numbers near the drift maximum, and their drift timescales are far shorter than the $\sim$1.7~Myr stellar age. The pebbles therefore migrated into the inner few tens of au long ago, consistent with our elevated gas-to-dust ratio of $\sim$270 and the gas-to-dust size ratio of $\gtrsim$3 (Section~\ref{sec:radial_drift}). Internal photoevaporation, by contrast, begins to open a gap only once the accretion rate falls below the photoevaporative wind rate \citep[e.g.,][]{Alexander2014, Owen2016}. The measured accretion rate of HD~179218, $\sim$$10^{-8}\,\msun\,{\rm yr}^{-1}$ \citep[][]{Mendigutia2017}, is at or below the wind rates of $\sim$$10^{-8}$--$10^{-7}\,\msun\,{\rm yr}^{-1}$ predicted for luminous intermediate-mass stars \citep[][]{Gorti2009, Komaki2021}, placing the system in the regime where clearing is expected to be underway---consistent with the partial ($\delta = 0.15$), rather than complete, gas depletion we infer. Crucially, photoevaporative winds entrain only grains smaller than a few $\mu$m \citep[][]{Franz2020}: the millimeter pebbles are left behind on their orbits as the gas is removed from beneath them, and as the local gas density drops their coupling to the gas weakens further, stalling any subsequent drift. This scenario thus accounts simultaneously for the compact dust interior to the gas cavity, the absence of a pressure-trap ring at the cavity edge, the partial gas depletion, and the elevated gas-to-dust ratio.

The main alternative---a giant planet at $\sim$30--50~au carving the gas cavity---predicts a millimeter dust ring at the outer cavity edge, where drifting pebbles collect at the pressure bump; such rings are the hallmark of planet-carved transition disks. The absence of a detected ring at 50~au argues against this scenario, although a faint ring below our sensitivity, or a configuration in which small grains filter through the gap and re-grow in the inner disk, cannot be fully excluded. Deeper, higher-resolution continuum imaging and searches for kinematic planetary signatures would distinguish the two scenarios. Finally, we consider two effects that could mimic an inner deficit in the line emission without gas depletion. Over-subtraction of the bright, compact 3~mm continuum would require the continuum to be optically thick, but the measured spectral index of $\alpha \approx 2.5$ (Section~\ref{sec:radial_drift}) is well above the optically thick blackbody value of $\alpha = 2$, indicating largely optically thin emission; we therefore consider this effect unlikely. Shadowing of the 10--50~au region by the $\sim$10~au inner disk \citep[][]{Fedele2008, Kokoulina2021} remains possible in principle, but cannot readily explain a depletion extending to 50~au, well beyond the continuum extent.

\subsection{Comparison to Other Herbig Disks and Implications for CO as a Gas Mass Tracer}
\label{sec:herbig_comparison}

To place HD~179218 in context, we compare its millimeter flux with the ALMA Herbig disk survey of \citet{Stapper2022}. We choose to use continuum fluxes rather than inferred dust masses, which depend on opacity and temperature assumptions. Scaled to 150~pc, our 1.3~mm flux is $31~\mathrm{mJy} \times (266/150)^2 \approx 98$~mJy, versus a sample median of $\approx$120~mJy (normalizing the Band~6 fluxes in the same way, with the \citet{Vioque2018} distances adopted by that work). HD~179218 is thus comparable to, though slightly below, the median and an order of magnitude below the brightest disks (e.g., HD~97048, HD~142527). Its dust-disk radius ($\sim$20--40~au) is at the low end of the resolved Herbig disks ($R_{\rm dust,68\%} \approx 35$--315~au, median $\approx$71~au), consistent with the substantial dust evolution discussed in Section~\ref{sec:radial_drift}.

A key distinction of our analysis is the explicit combination of continuum and CO isotopologue modeling to measure both dust and gas masses independently. While \citet{Stapper2022} relied on continuum emission alone, our approach allows us to directly constrain the gas-to-dust ratio and, because of the negligible CO freeze-out in HD~179218, to do so without the usual CO-depletion caveats.

Our measurement therefore provides a benchmark: when CO freeze-out is absent, CO isotopologues recover a substantial gas reservoir ($\mgas \sim 8\times10^{-3}\,\msun$) without the severe depletion inferred in many cooler disks. This contrasts with the much lower gas masses historically derived from CO emission in T~Tauri disks, where freeze-out and chemical processing reduce the observable CO. Those low values, however, have recently been shown to be largely an artifact of CO underabundance rather than genuinely gas-poor disks: using the freeze-out-insensitive N$_2$H$^+$ tracer in combination with CO, \citet{Trapman2025b} find median gas-to-dust ratios of $\sim$40--120 across the AGE-PRO sample of T~Tauri disks, with bulk CO abundances roughly an order of magnitude below the ISM value. Consistently, \citet{Trapman2025} find from a comparison of N$_2$H$^+$- and CO-based masses that CO underestimates the gas mass far less severely in the warmer disks around Herbig stars than in cooler T~Tauri disks. HD~179218, in which freeze-out is absent in our best-fitting model, provides an independent confirmation of the same conclusion from the opposite direction: when CO cannot be depleted by freeze-out, it recovers a physically reasonable, ISM-exceeding gas-to-dust ratio. Together, these results indicate that the dominant systematic in CO-based gas masses is the extent of CO freeze-out set by the disk thermal structure, and that CO isotopologues can trace the gas free of this bias in disks warm enough that freeze-out cannot operate.

\section{Conclusions}
\label{sec:conclusions}

We have measured the gas and dust masses of the protoplanetary disk around the Herbig star HD~179218 using NOEMA and ALMA observations of six CO isotopologue lines---$^{13}$CO, C$^{18}$O, and C$^{17}$O, each observed in the $J=2$--1 and $J=1$--0 transitions; the observed but undetected $^{13}$C$^{18}$O $J$=2--1 line is not counted among the six. The high stellar luminosity of HD~179218 ($\sim$100~$L_\odot$) ensures that CO remains in the gas phase throughout the disk, making this system a rare, clean test of CO-based gas mass measurements. Our main conclusions are:

\begin{enumerate}
    \item The dust mass, constrained from SED fitting with \texttt{RADMC-3D}, is $\mdust \approx 3 \times 10^{-5}\,\msun$ for a maximum grain size of 1~mm. Its millimeter flux is comparable to the median of the Herbig disk population (Section~\ref{sec:herbig_comparison}).

    \item The gas-disk characteristic radius is $\rc \approx 90$~au, constrained by the optically thin C$^{18}$O and C$^{17}$O emission. Because the $^{13}$CO emission is optically thick and primarily traces the temperature of the emitting surface rather than the gas column density, it is excluded from the fit. The models nevertheless overpredict the $^{13}$CO emission, potentially due to an overestimated gas temperature in its emitting layer.

    \item The gas mass, measured from the optically thin C$^{18}$O and C$^{17}$O lines using DALI thermochemical models, is $\mgas \approx 8 \times 10^{-3}\,\msun$, yielding a gas-to-dust mass ratio of $\sim$270.

    \item The absence of CO freeze-out, confirmed by the DALI thermal structure, means that this gas mass measurement is not subject to the CO depletion that can bias CO-based mass estimates in cooler disks. Isotope-selective photodissociation, a separate process in the disk surface layers, is still present and is treated self-consistently in our DALI models. The elevated gas-to-dust ratio is therefore not an artifact of CO depletion, although its absolute value depends on the dust opacity.

    \item Several lines of evidence are consistent with substantial inward drift of millimeter dust. Its dust disk radius ($\sim$20--40~au) is much more compact than the gas disk traced by $^{13}$CO ($R_{68} \approx 120$~au), directly showing differential dust and gas evolution. The elevated gas-to-dust ratio of $\sim$270 and the 1.3--3~mm spectral index ($\alpha \approx 2.5$, indicating grains grown to at least millimeter sizes) are also consistent with this picture, although the gas-to-dust ratio is subject to dust-mass systematics.

    \item The disk exhibits an unusual structure: a gas cavity at 50~au ($\delta = 0.15$) with the compact millimeter dust located \textit{inside} it, and no dust ring at the cavity edge---the inverse of a typical transition disk. We propose that this records a two-step evolution: millimeter pebbles first drifted rapidly into the inner disk, and internal photoevaporation subsequently began removing the inner gas once the accretion rate ($\sim$$10^{-8}\,\msun\,{\rm yr}^{-1}$) fell to the level of the expected photoevaporative wind rates. Because winds entrain only micron-sized grains, the pebbles are left behind as the gas is cleared from beneath them, naturally producing a gas cavity without a corresponding dust cavity.
\end{enumerate}

These results demonstrate that, in disks where the CO snowline lies beyond the disk edge, CO isotopologues trace the gas-phase carbon at its full abundance and yield gas-to-dust ratios that are not biased by freeze-out, even if the gas-to-dust ratio remains uncertain at the factor-of-a-few level set by the dust opacity. The contrast with the low CO-based masses measured in cooler disks strengthens the case that CO depletion---driven by freeze-out and subsequent chemical or physical processing---is a major contributor to the discrepancy between CO-based and independent gas mass estimates.

\begin{acknowledgments}
N.W. acknowledges support from an undergraduate research fellowship from the Wisconsin Space Grant Consortium and the Doherty Undergraduate Award from the Department of Astronomy at the University of Wisconsin--Madison. This paper makes use of the following ALMA data: ADS/JAO.ALMA\#2021.1.0709.S. ALMA is a partnership of ESO (representing its member states), NSF (USA) and NINS (Japan), together with NRC (Canada), MOST and ASIAA (Taiwan), and KASI (Republic of Korea), in cooperation with the Republic of Chile. The Joint ALMA Observatory is operated by ESO, AUI/NRAO and NAOJ. The National Radio Astronomy Observatory is a facility of the National Science Foundation operated under cooperative agreement by Associated Universities, Inc. This work is based on observations carried out under project number S20AR with the IRAM NOEMA Interferometer. IRAM is supported by INSU/CNRS (France), MPG (Germany) and IGN (Spain).
\end{acknowledgments}

\facilities{ALMA, NOEMA}


\software{
  astropy \citep{Astropy_2022,Astropy_2018, Astropy_2013},
  CASA \citep{CASA2022},
  GILDAS \citep{GILDAS2013},
  RADMC-3D \citep{Dullemond2012},
  DALI \citep{bruderer2012, bruderer2013},
  matplotlib \citep{Hunter2007}
}
\bibliographystyle{aasjournal}
\bibliography{references}

\begin{appendix}

\section{Excluded Lines: C$^{17}$O $J$=1--0 Radial Profile and $^{13}$CO $J$=2--1 Spectrum}
\label{app:excluded_lines}

Figures~\ref{fig:C17O_best_radial} and \ref{fig:13CO_best_spectra} show the best-fitting model compared to the two transitions excluded from our $\chi^2$ (Section~\ref{sec:DALI_method}). The C$^{17}$O $J$=1--0 radial profile (Figure~\ref{fig:C17O_best_radial}) is contaminated by the blending of its three hyperfine components in the inner region, which produces additional observed emission that is not included in the modeled profile and therefore causes the model to underpredict the observed intensity in this region. The contaminated radial range is indicated by the shaded region in Figure~\ref{fig:C17O_best_radial}. The $^{13}$CO $J$=2--1 spectrum (Figure~\ref{fig:13CO_best_spectra}) is overpredicted by the model, consistent with the systematic $^{13}$CO overprediction discussed in Section~\ref{sec:disk_size}.

\begin{figure*}[t]
    \centering
    \includegraphics[width=\textwidth]{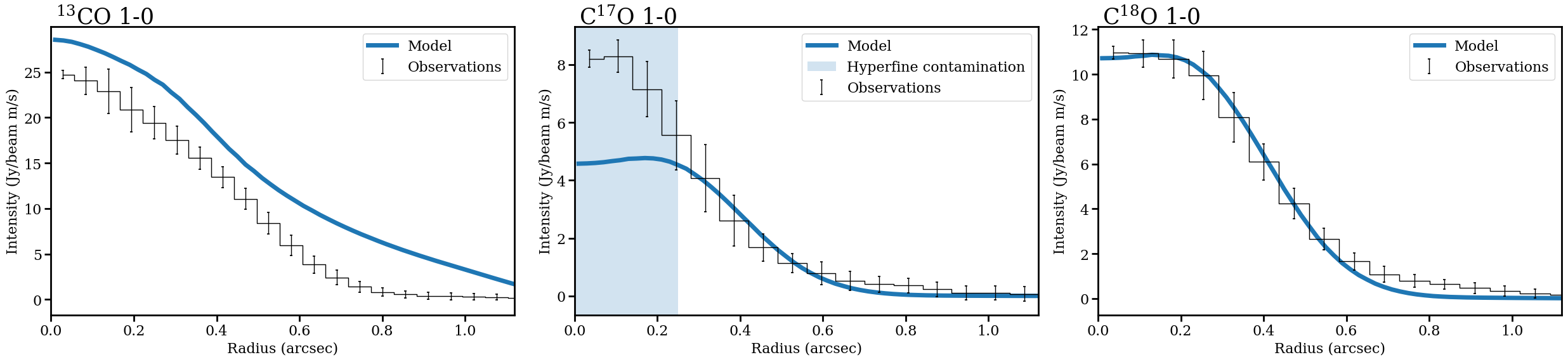}
    \caption{Radial profiles of the best-fitting DALI model (blue) compared to ALMA observations (black) for the $^{13}$CO, C$^{17}$O, and C$^{18}$O $J$=1--0 lines. The C$^{17}$O model underpredicts the observations due to contamination from three hyperfine components.}
    \label{fig:C17O_best_radial}
\end{figure*}

\begin{figure*}[t]
    \centering
    \includegraphics[width=\textwidth]{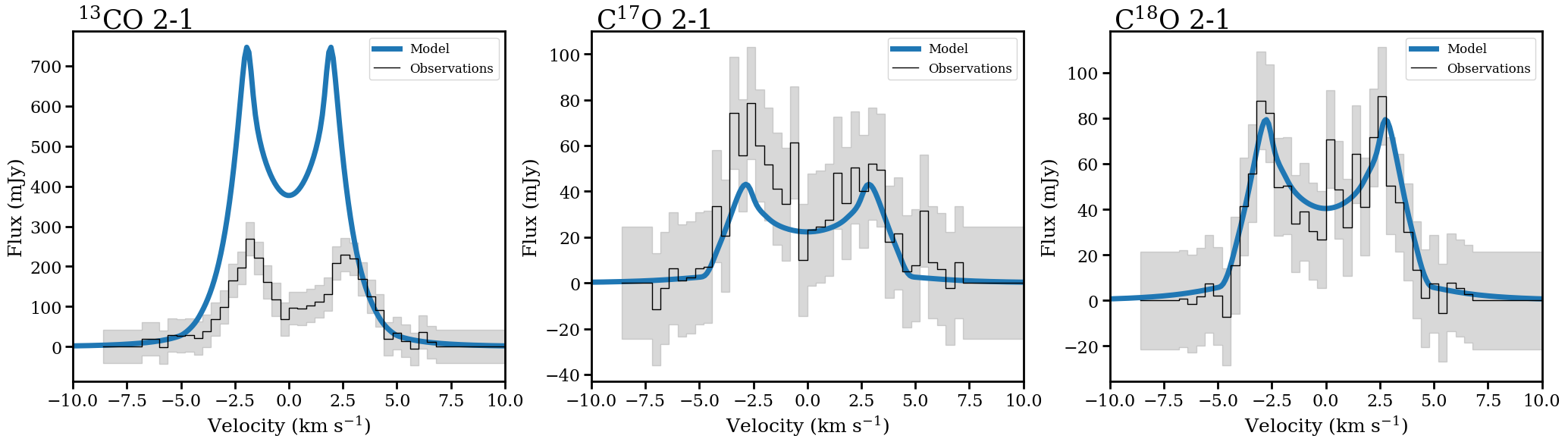}
    \caption{NOEMA $J$=2--1 spectra (black) compared to the best-fitting DALI model (blue) for $^{13}$CO, C$^{18}$O, and C$^{17}$O.}
    \label{fig:13CO_best_spectra}
\end{figure*}

\end{appendix}

\end{document}